\documentclass[floatfix,aps,twocolumn,nofootinbib,superscriptaddress]{revtex4}

\usepackage{graphicx}
\usepackage{epsfig}

\usepackage{amsmath}
\usepackage{amsfonts}
\usepackage{amssymb}
\usepackage{bm}
\usepackage{mathtools}

\usepackage{hyperref}
\usepackage{dsfont}
\usepackage{lipsum}

\usepackage{tikz}
\usepackage{circuitikz}
\usetikzlibrary{arrows, shapes}
\usepackage{pgfplots}
\pgfplotsset{compat=1.14}

\usepackage{pdfpages}

\newcommand{\mean}[1]{\left\langle #1 \right\rangle}
\newcommand{\meann}[1]{\langle #1 \rangle}

\begin{document}
\author{Nahuel Freitas}
\affiliation{Complex Systems and Statistical Mechanics, Department of Physics and Materials Science,
University of Luxembourg, L-1511 Luxembourg, Luxembourg}

\author{Massimiliano Esposito}
\affiliation{Complex Systems and Statistical Mechanics, Department of Physics and Materials Science,
University of Luxembourg, L-1511 Luxembourg, Luxembourg}

\title{A Maxwell demon that can work at macroscopic scales}

\date{\today}
\begin{abstract}
Maxwell's demons work by rectifying thermal fluctuations. They are not expected to function at macroscopic scales where fluctuations become negligible and dynamics become deterministic. 
We propose an electronic implementation of an autonomous Maxwell's demon that indeed stops working in the regular macroscopic limit as the dynamics becomes deterministic. However, we find that if the power supplied to the demon is scaled up appropriately, the deterministic limit is avoided and the demon continues to work. The price to pay is a decreasing thermodynamic efficiency. Our work suggests that novel strategies may be found in nonequilibrium settings to bring to the macroscale non-trivial effects so far only observed at microscopic scales.
\end{abstract}
\maketitle

\emph{Introduction}. A famous thought experiment proposed by James C. Maxwell in 1867 introduced the idea of an agent with the ability to perceive the velocity of individual molecules in a gas, and open or close molecular valves based on its observations \cite{leff2002}. Such agent, or Maxwell's Demon (MD), could in that way create gradients of temperature or pressure without expending any work, in apparent contradiction with the second law of thermodynamics. Indeed, the intention of Maxwell was to illustrate the statistical nature of this law, that according to him was only valid at macroscopic scales where we have `no power of handling or perceiving separate molecules' \cite{leff2002}. Later developments proved Maxwell to be partially wrong: the second law is statistical, but it is also valid at microscopic scales.
Thus, a MD must necessarily dissipate energy and produce entropy, in such a way that the second law is still respected at a global level of description that includes the MD. Nevertheless, the resolution of the issue led to a revolution in our understanding of thermodynamics, that unveiled its deep connection with the abstract notion of information \cite{bennett1982, parrondo2015}. In modern terms, MDs or `information engines' are understood as active feedback control loops acting on a fluctuating system, and the second law  has been successfully extended to those settings \cite{Sagawa2008, Sagawa2009, Cao2009, Sagawa2012, esposito2012, barato2014, horowitz2014, hartich2014}. 

A first experimental realization of a MD was achieved 15 years ago in a molecular system powered by light \cite{Serreli2007}.
After that, MDs were implemented in a great variety of systems such as trapped atoms \cite{Price2008, Raizen2009,
Bannerman2009, Kumar2018}, colloidal particles \cite{Toyabe2010, saha2021}, single electron circuits \cite{Koski2014Sep, Koski2014Jul, Koski2015, Chida2017}, nuclear spins \cite{Camati2016}, electro-photonic systems \cite{Vidrighin2016}, superconducting qubits \cite{Cottet2017, Masuyama2018, Naghiloo2018}, DNA hairpin pulling experiments \cite{Ribezzi-Crivellari2019}, and cavity QED setups \cite{Najera-Santos2020}. All those implementations share a common feature with the `very observant and neat-fingered' agent Maxwell imagined: they work at the microscopic level. They are able to detect fluctuations at the level of individual particles, molecules, electrons, or photons, depending on the setup. This is only natural, given that the task of a MD is to rectify thermal fluctuations, and those fluctuations are negligible at macroscopic scales (as is known from equilibrium thermodynamics, they decrease as $1/\sqrt{\Omega}$, where $\Omega$ is an adimensional scale parameter \cite{pathria2021}).
One exception are systems with continuous phase transitions, during which the  symmetry of a system is spontaneously broken in favour of one of several metastable states that are macroscopically distinct. Since the final metastable state is selected by thermal fluctuations, continuous phase transitions can be employed to transfer information from the micro to the macro level \cite{roldan2014}, and macroscopic MDs can be devised based on this principle \cite{parrondo2001}. 


In this work, we propose an alternative way to construct macroscopic MDs by realizing that thermodynamic resources can be employed to actively amplify thermal fluctuations from the micro to the macro level. Specifically, we propose a practical electronic implementation of an \emph{autonomous} MD that, to the best of our knowledge, is the first one to have a well defined macroscopic limit. In addition, it only involves regular CMOS technology, the one powering modern computers.
By only changing the width of the conduction channel of the MOS transistors, the proposed design can work from the single-electron regime
all the way up to macroscopic regimes of operation. We find that when all additional parameters are fixed, the rectification effects associated to the action of the MD disappear above a finite scale $\Omega^*$. However, as we show analytically and numerically, when the power available to the MD is appropriately scaled, the rectification effects survive in the macroscopic limit. In that case, the price to pay is a decreasing thermodynamic efficiency, that we show scales as $\eta \sim 1/\Omega^2\log(\Omega)$.

\begin{figure*}
\centering
\includegraphics[width=\textwidth]{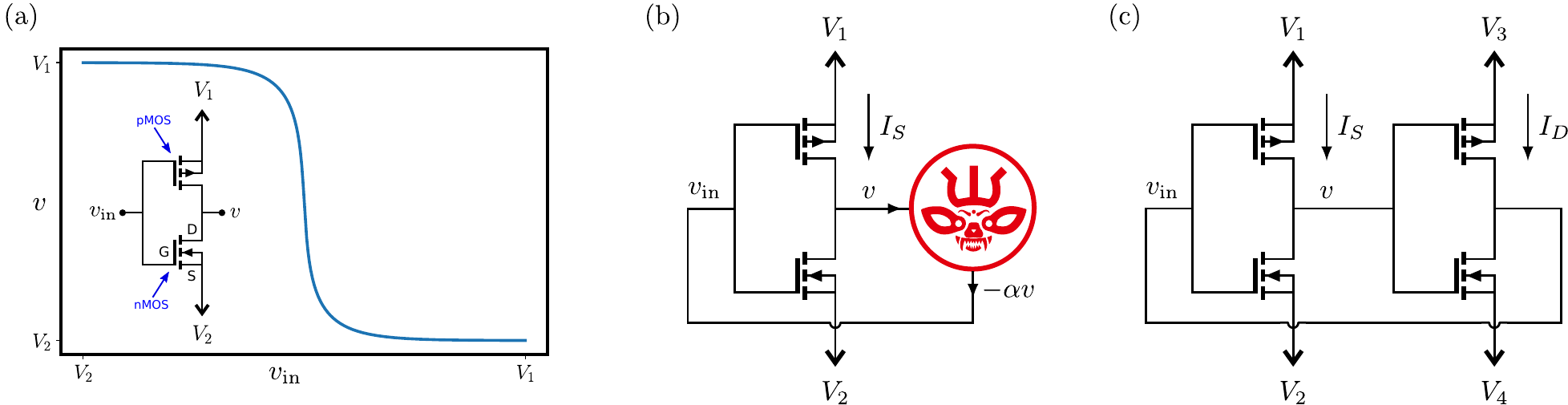}
\caption{(a) Typical deterministic transfer function for the common implementation of a sub-threshold CMOS inverter. (b) An external agent measures the fluctuating output voltage $v$ of a CMOS inverter and adjusts the input voltage as $v_\text{in} = -\alpha v$, with $\alpha > 0$.
(c) A similar feedback protocol is implemented with an additional CMOS inverter.}
\label{fig:circuit_join}
\end{figure*}

\emph{A full-CMOS Maxwell's Demon}. Our design is based on the common implementation of a CMOS inverter, or NOT gate, shown in Figure \ref{fig:circuit_join}-(a). It is composed of a nMOS transistor and a
pMOS transistor operating in the sub-threshold regime \cite{wang2006}.
In a three-terminal nMOS (pMOS) transistor, electrical conduction between drain (D) and
source (S) terminals is exponentially enhanced (reduced) for increasing gate (G) to source voltage. 
As a consequence, when a voltage bias is applied by
connecting the drain terminals to voltage sources $V_1$ and $V_2 = - V_1$, the
input-output transfer function of the inverter has the typical shape shown in
Figure \ref{fig:circuit_join}-(a). For positive values of the input, conduction through the nMOS
transistor dominates and the output voltage approaches $V_2$.  The situation is
reversed for negative input voltages, and the output voltage approaches $V_1$.
If no bias is applied, $V_1=V_2=0$, the circuit attains thermal equilibrium 
and the steady state fluctuations in the output voltage $v$ are given by
$P_\text{eq}(v) \propto e^{-\beta U(v)}$, where $\beta=1/k_b T$ is the inverse
temperature and $U(v) = v^2/2C$ the electrostatic energy. 
$C$ is the output capacitance of the inverter related to the physical scale of the transistors. 
Then, the mean output voltage is $\mean{v}=0$, with a standard deviation $\sigma = \sqrt{k_bT/C}$. 
Let us consider the following feedback protocol, carried on by some external agent or device 
(see Figure \ref{fig:circuit_join}-(b)). At time $t$, the output voltage $v$ is measured and
a voltage $v_\text{in} = -\alpha v$ is applied to the input, with $\alpha >0$.
If a positive fluctuation $v>0$ is observed at time $t$, at subsequent
times conduction through the upper pMOS transistor will be enhanced with
respect to conduction through the nMOS transistor (since the input voltage at
the gates will be negative). Therefore, the excess charge will be most likely
dissipated through the pMOS transistor, generating a net upward current. 
In a similar way, when a negative fluctuation $v<0$ is observed, it will most probably be compensated 
by conduction events through the bottom nMOS transistor, also generating a net upward current. 
Thus, by the repeated application of the feedback protocol, charge can be made to flow in the upward
direction. We can say that by controlling the input voltage, the demon opens or closes electronic 
doors forcing the electrons to flow in one direction. 
A net upward electric current could also be observed even if a small downward bias 
$\Delta V_S = V_1 - V_2 > 0$ is applied. In that case, the entropy production rate 
$\dot \Sigma_S = \Delta V_S \mean{I_S}/T$ is negative ($\mean{I_S}$ is the average 
steady state electric current through the inverter, and therefore $\Delta V_S \mean{I_S} < 0$ 
is the rate of heat dissipation). 
According to the second law of thermodynamics, and our modern understanding of Maxwell's demons, 
that negative entropy production must be compensated by a positive entropy production in the device 
implementing the feedback control protocol (the MD). To study this, one must consider an explicit mechanism for the control device and turn to an autonomous picture that includes the MD. We do so by noticing that the operation $v_\text{in} \to -\alpha v$ can be approximately implemented by an additional CMOS inverter, as shown in Figure \ref{fig:circuit_join}-(c). This additional inverter is powered by a voltage bias $\Delta V_D = V_3-V_4$, which 
generates a current $I_D$. 
If we break the left/right symmetry of the full circuit by choosing $\Delta V_D > \Delta V_S$, we 
can consider that the second inverter (the demon) continuously monitors the output of the first inverter, 
and adjusts its input accordingly. Whenever the demon produces an average current 
$\mean{I_S}$ against the bias $\Delta V_S$, its efficiency is 
\begin{equation}
    \eta = -\dot \Sigma_S/\dot \Sigma_D \geq 0,
\end{equation}
where $\Sigma_{D} = \Delta V_{D} \mean{I_{D}}/T$ is the rate of entropy production at the demon side. 
The total entropy production is $\dot \Sigma = \Sigma_S + \Sigma_D \geq 0$ and implies that $\eta \leq 1$.

We show next how to model the full circuit in Figure \ref{fig:circuit_join}-(c) both at the deterministic and stochastic levels. Importantly, and at variance with the idealized feedback protocol described above, the stochastic description fully accounts for the intrinsic thermal noise at the demon side, and the delay between measurement and feedback.

\begin{figure*}[t]
\centering
\includegraphics[scale=.395]{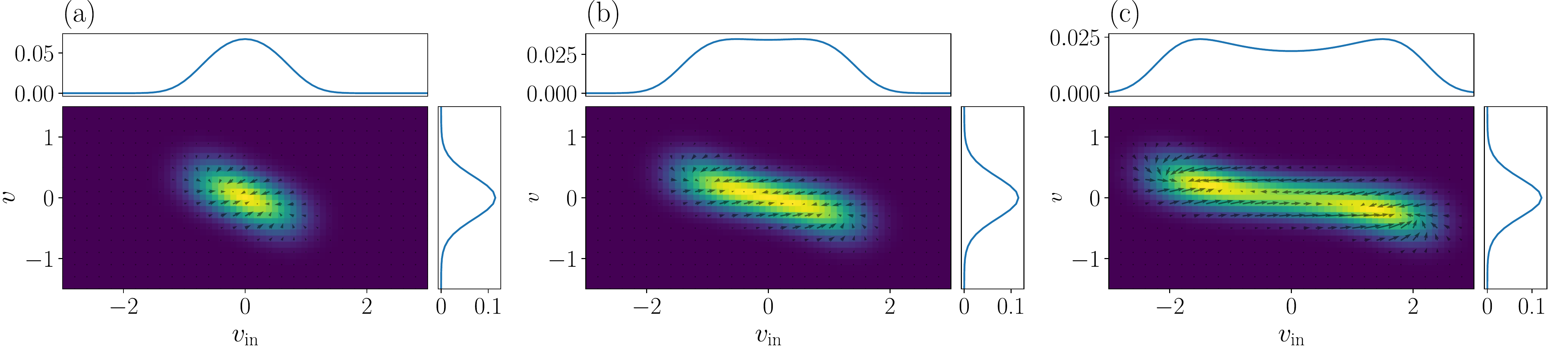}
\caption{Histograms of the steady state distribution $P_\text{ss}(\bm{v})$ for $v_e = 0.1$, $\Delta V_S = 0.4$ and: (a) $\Delta V_D=2$ ($\alpha^2 < 1$, monostable phase), (b) $\Delta V_D\simeq 3.415$ ($\alpha^2 =1$, critical point), 
and (c) $\Delta V_D = 5$ ($\alpha^2 >1$, bistable phase). Lateral and upper plots show the partial distributions for $v$ and $v_\text{in}$, respectively. The arrows indicate the direction and magnitude of the flow $\bm{u}(\bm{v}) = \sum_\rho \bm{\Delta}_\rho \: \omega_\rho(\bm{v}) P_\text{ss}(\bm{v})$ at each point in the state space.}
\label{fig:histograms}
\end{figure*}

\emph{Deterministic description.}
All voltages will be expressed in units of the thermal voltage $V_T = k_bT/q_e$, where $q_e$ is the absolute value of the electron charge. We consider powering voltages $V_1 = -V_2 = \Delta V_S/2$ and $V_3 = -V_4 = \Delta V_D/2$. 
The deterministic equations of motion for the voltages $v$ and $v_\text{in}$ -- the two degrees of freedom in the circuit of Figure \ref{fig:circuit_join}-(b) -- are
\begin{equation}
\begin{split}
    C\: d_t v &= I_\text{p}(v, v_\text{in} ; \Delta V_S) - I_\text{n}(v, v_\text{in} ; \Delta V_S) \\
    C\: d_t v_\text{in} &= I_\text{p}(v_\text{in}, v ; \Delta V_D) - I_\text{n}(v_\text{in}, v ; \Delta V_D),
    \label{eq:det_dyn}
\end{split}
\end{equation}
where $I_\text{(n/p)}(v, v_\text{in}; \Delta V)$ are the electric currents through the (n/p)MOS transistors for given voltages $v$ and $v_\text{in}$.  In the sub-threshold regime of operation, the pMOS current is given by \cite{wang2006}:
\begin{equation}
 I_\text{p}(v, v_\text{in};\Delta V)=I_0e^{(\Delta V/2-v_\text{in}-V_\text{th})/n}\left(1-e^{-(\Delta V/2-v)}\right),
 \label{eq:iv_pmos}
\end{equation}
where $I_0$ (the specific current), $V_\text{th}$ (the threshold voltage) and $n\geq 1$ (the slope factor), are phenomenological parameters characterizing the transistor. Assuming, for simplicity, that these parameters are the same for all transistors, the nMOS current is given by $I_\text{n}(v, v_\text{in};\Delta V) = I_\text{p}(-v, -v_\text{in};\Delta V)$. Close to thermal equilibrium (i.e. for low biases $\Delta V_S$ and $\Delta V_D$), the previous deterministic dynamics has the unique fixed point $v = v_\text{in} = 0$, but the system becomes bistable when $\alpha^2 \equiv \alpha_S \alpha_D > 1$, where $\alpha_{S/D} \equiv (1-e^{\Delta V_{S/D}/2})/n$. This is easily understood by noticing that, for small inputs, each inverter works as a linear amplifier with gain $\alpha_{S/D}$. The bistability can be exploited to store the value of a bit. Indeed, for symmetric powering $\Delta V_S = \Delta V_D = \Delta V$ the circuit in Figure \ref{fig:circuit_join} is the typical CMOS implementation of SRAM memory cells. Note that the fixed points of Eqs. \eqref{eq:det_dyn} always satisfy $V_1>v>V_2$ and $V_3>v_\text{in}>V_4$, implying that all electric currents are positive and thus that no rectification is possible at the deterministic level.  

\emph{Stochastic description.}
Conduction through the MOS transistors is not deterministic. The number of charges transported through the channel of a MOS transistor during a given time interval is a stochastic quantity that, in the sub-threshold regime of operation, displays shot noise \cite{sarpeshkar1993}. This can be described by an effective model that assigns two Poisson processes to every transistor, corresponding to elementary conduction events in the forward and backward directions. The two Poisson rates can be computed from the I-V curve characterization of the device and the requirement of thermodynamic consistency \cite{freitas2021}. The dynamics becomes a Markov jump process in the space $\{\bm{v}\}$ of possible circuit states and the probability distribution $P_t(\bm{v})$ over those states evolves according to the Markovian master equation
\begin{equation}
d_t P_t(\bm{v}) = \sum_{\rho}\lambda_\rho(\bm{v}-\bm{\Delta}_\rho v_e)P_t(\bm{v}-\bm{\Delta}_\rho v_e)-\lambda_\rho(\bm{v})P_t(\bm{v}),
\label{eq:master_eq_auto}
\end{equation}
where $\bm{v} = (v, v_\text{in})$ is a vector with the two degrees of freedom of the circuit, 
$\lambda_\rho(\bm{v})$ are the Poisson jump rates assigned to the different transistors (that are indexed by $\rho)$, and the vectors $\bm{\Delta}_\rho$ encode the change in $\bm{v}$ for each jump. For example, for forward transitions through the pMOS and nMOS transistors of the first inverter, we have $\bm{\Delta}_\rho = (1,0)$ and $\bm{\Delta}_\rho = (-1,0)$, respectively. The elementary voltage $v_e \equiv q_e/C$ is the magnitude by which the voltages change in each jump. For example, the two Poisson rates for the pMOS I-V curve in Eq. \eqref{eq:iv_pmos} are:
\begin{equation}
\begin{split}
\lambda_+^\text{p}(v,v_\text{in};\Delta V) &= \tau_0^{-1} \: e^{(\Delta V/2-v_\text{in})/n} \qquad\\
\lambda_-^\text{p}(v,v_\text{in};\Delta V) &= \lambda_+^p(v,v_\text{in};\Delta V) \: e^{-(\Delta V/2-v)} \: e^{-v_e/2},
\end{split}
\label{eq:mos_rates}
\end{equation}
where $\tau_0 = (I_0 e^{-V_\text{th}/n}/q_e)^{-1}$ sets the system timescale. 

\begin{figure}[t]
\centering
\includegraphics[scale=.50]{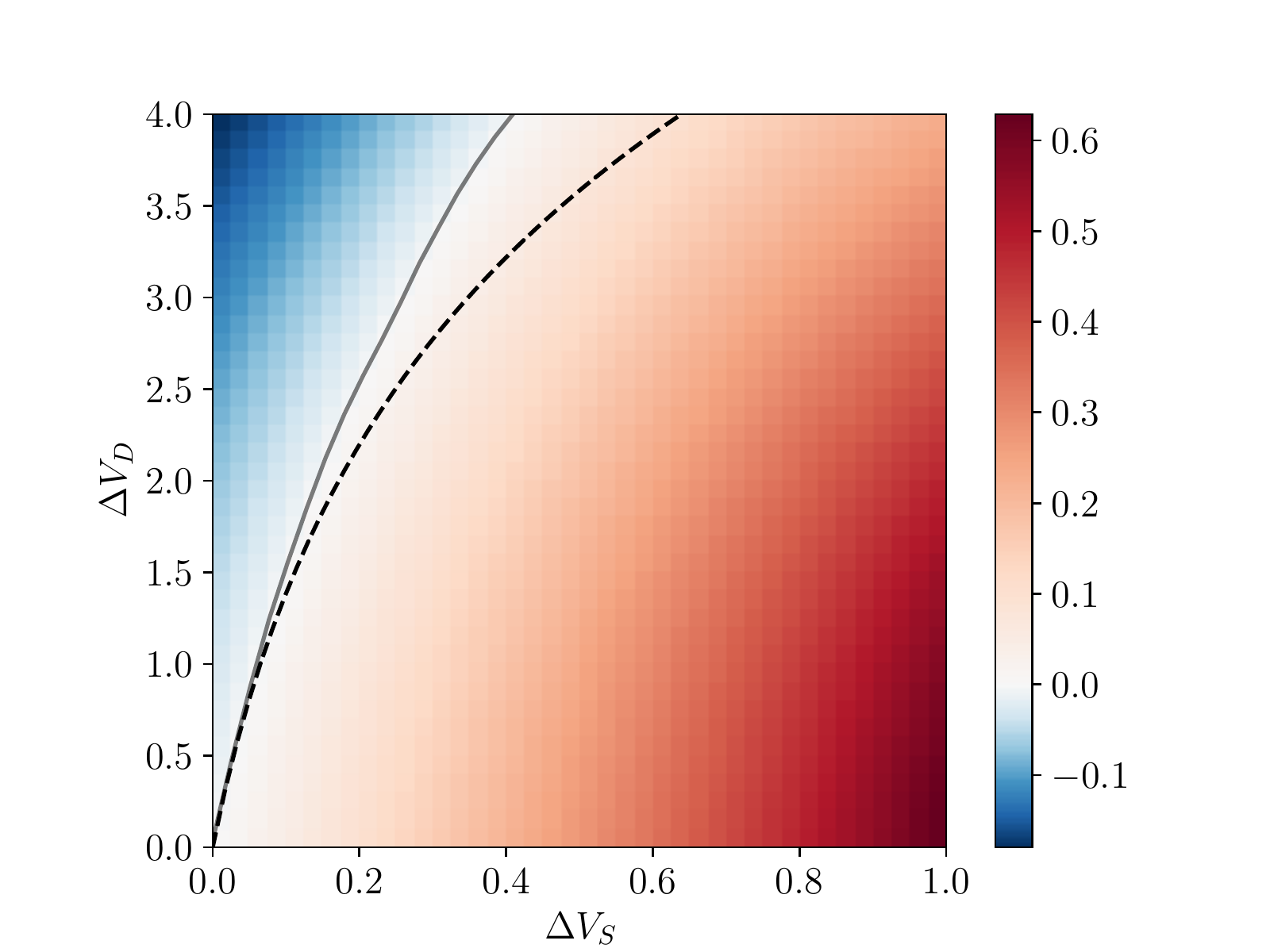}
\caption{Average electric current $\mean{I_S}$ (in units of $q_e/\tau_0$) at steady state for the circuit of Figure \ref{fig:circuit_join}-(c) as a function of the powering voltage biases $\Delta V_S$ and $\Delta V_D$ ($v_e=0.1$, $n=1$). The solid gray line marks the boundary between the $\mean{I_S}<0$ (blue) and $\mean{I_S}>0$ (red) regions, and the dashed line corresponds to the approximate analytical result of Eq. \eqref{eq:min_vd}.}
\label{fig:auto_current}
\end{figure}

The steady state $P_\text{ss}(\bm{v})$ of Eq. \eqref{eq:master_eq_auto} can be obtained by spectral methods or by generating stochastic trajectories via the Gillespie algorithm. 
From it, one can compute the average electric currents through the different transistors in the circuit.
In Figure \ref{fig:histograms}, we plot $P_\text{ss}(\bm{v})$ for $v_e = 0.1$, $\Delta V_S = 0.4$, and three different values of $\Delta V_D$, corresponding to the monostable phase ($\alpha^2 <1$), the critical point ($\alpha^2=1$), and the bistable phase ($\alpha^2 >1$). In Figure \ref{fig:auto_current} we show the average current through the first inverter for $v_e=0.1$ and different values of $\Delta V_S$ and $\Delta V_D$. Crucially, we now see that the current can indeed be reversed for sufficiently high values of $V_D$.

\emph{Macroscopic limit.} We now study the stochastic behaviour of the circuit as the physical dimensions of the transistors are increased. There are two relevant length scales: the width $W$ and the length $L$ of the conduction channel in each transistor. For fixed $L$, both the parameter $I_0$ appearing in Eq. \eqref{eq:iv_pmos} (that enter the Poisson rates through the timescale $\tau_0$) and the capacitance $C$ are proportional to $W$. Thus, the Poisson rates $\lambda_\pm^\text{n/p}(\bm{v})$ increase as $W$, while the elementary voltage $v_e = q_e/C$ decreases as $W^{-1}$. We therefore consider $v_e^{-1} \propto W$ as the scale parameter.  
From the previous observations, it follows that the probability distribution $P_t(\bm{v})$ satisfies a large deviations (LD) principle in the limit $v_e \to 0$. This means that fluctuations away from the deterministic behaviour are exponentially suppressed in $v_e^{-1}$. Mathematically, the LD principle implies the existence of the limit
$f(\bm{v},t) = \lim_{v_e \to 0} -v_e \log(P_t(\bm{v}))$, also expressed as:
\begin{equation}
P_t(\bm{v}) \asymp e^{-(f(\bm{v},t) + o(v_e))/v_e},
\label{eq:lda}
\end{equation}
where $f(\bm{v},t)$ is the rate function (i.e. the rate at which the probability of fluctuation $v$ decreases with $v_e^{-1}$). Indeed, by plugging the previous ansatz in the master equation \eqref{eq:master_eq_auto} and keeping only 
the dominant terms in $v_e \to 0$, we find that the rate function must satisfy:
\begin{eqnarray}
 d_t f(\bm{v},t) =\sum_\rho \omega_\rho(\bm{v})\left[1-e^{\bm{\Delta}_\rho\cdot \nabla f(\bm{v},t)}\right],
 \label{eq:rf_dyn}
\end{eqnarray}
where $\omega_\rho(\bm{v}) = \lim_{v_e \to 0} v_e \lambda_\rho(\bm{v})$ are the scaled Poisson rates. The minima $\bm{v}(t)$ of the rate function $f(\bm{v}, t)$ at time $t$ are the most probable values. They evolve according to the closed deterministic dynamics of Eq. \eqref{eq:det_dyn} \cite{freitas2021LR}. This shows how the deterministic dynamics emerges from the stochastic one. 

Eq. \eqref{eq:rf_dyn} cannot be solved exactly, even at steady state. In order to make analytical progress, we perform a Gaussian approximation around $\bm{v}=0$, the most probable value in the monostable phase ($\alpha^2<1$). 
In this way we can compute the average steady state current through the first inverter to first non-trivial order in $v_e \to 0$ (see Supplementary Material). While its full expression is complicated, to first order in $\Delta V_S$ it reads:
 \begin{equation}
    \mean{I_S} \simeq \frac{q_e}{2\tau_0} \left[ \Delta V_S -v_e \left(e^{\Delta V_D/2}-1\right)\right].
    \label{eq:current_auto_linear}
\end{equation}
The first term corresponds to the deterministic current 
$I_S = q_e\Delta V_S/2\tau_0 + \mathcal{O}(\Delta V^2)$ which is incompatible with rectification (it is always positive for $\Delta V_S >0$). The second term describes the effect of the fluctuations and the feedback control, and it is always negative. Then, we see that the minimum value of $v_e$ for which rectification is possible (i.e. $\mean{I_S} <0$) is
$v_e^*  =\Delta V_S/\alpha_D$.
Thus, the rectification effects associated to the action of the MD disappear above a scale $1/v_e^*$, that increases with the amplification factor $\alpha_D$ of the MD, and decreases with the opposing bias $\Delta V_S$. Similarly, the minimum value of $\Delta V_D$ necessary for rectification 
is
\begin{equation}
    \Delta V_D^*  = 2\log(1+\Delta V_S/v_e).
    \label{eq:min_vd}
\end{equation}
The previous expression is plotted with a dashed line in Figure \ref{fig:auto_current}. The deviations from the exact $\mean{I_S} = 0$ boundary are due to non-Gaussian effects and the fact that Eq. \eqref{eq:current_auto_linear} is only valid for first order in $\Delta V_S$.

\begin{figure}
\centering
\includegraphics[scale=.55]{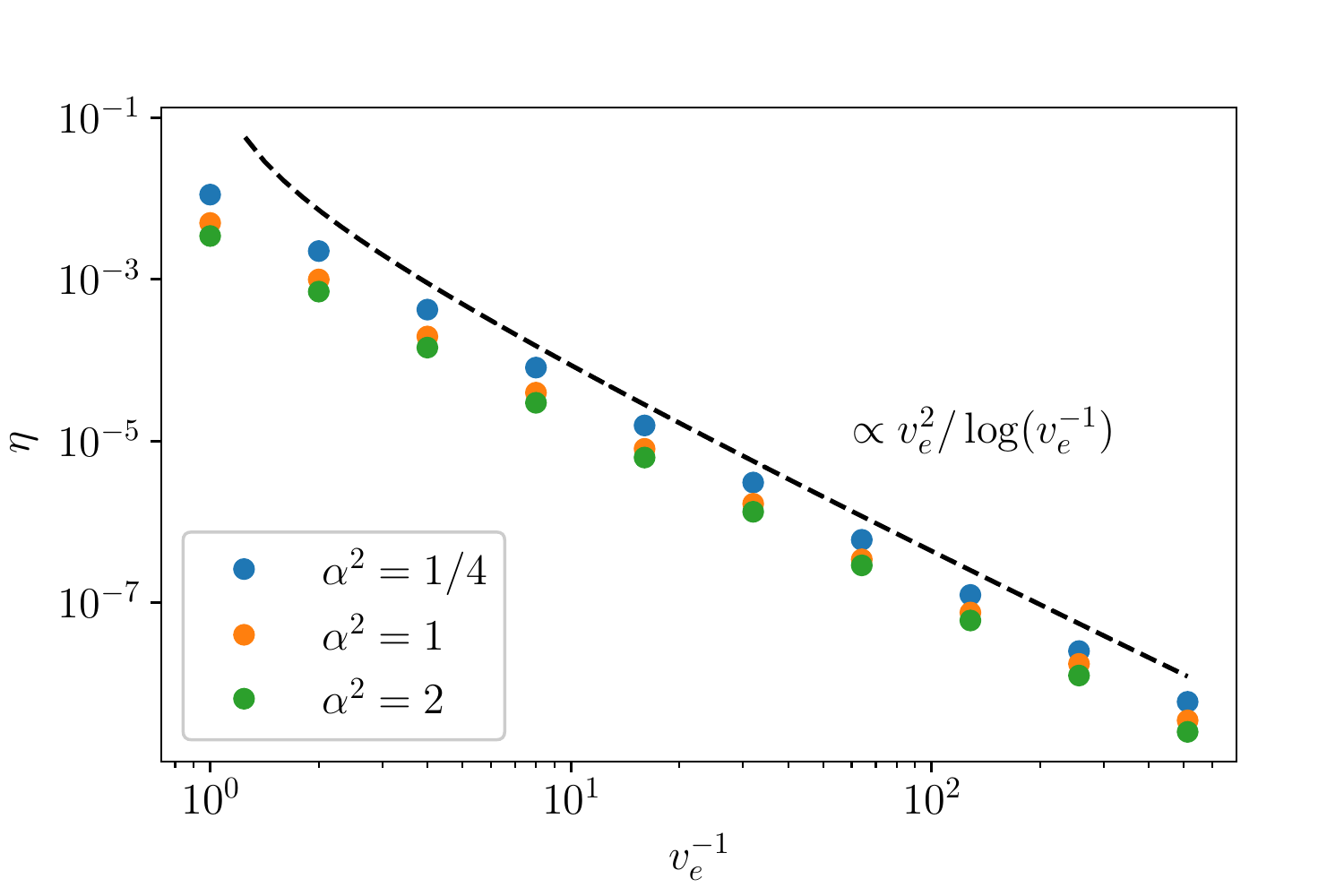}
\caption{Efficiency as a function of the scale parameter $v_e^{-1}$ with the scaling strategy $\Delta V_S = c v_e$ and $\Delta V_D = 2\log(1+2\alpha^2/cv_e)$ for $c=0.1$ and different values of $\alpha^2$. The results are obtained using the Gillespie algorithm. The dashed black line shows the scaling obtained from Eqs. \eqref{eq:scaling_system} and \eqref{eq:scaling_demon}.}
\label{fig:efficiency}
\end{figure}

\emph{Scaling analysis}. Although we did not assumed anything about the magnitude of $\Delta V_D$, the Gaussian approximation leading to Eq. \eqref{eq:current_auto_linear} is only valid for $\alpha^2 < 1$.
This imposes a maximum value for $\Delta V_D$ given any non-zero value of $\Delta V_S$. To take this into account, in the following we will consider a constant value of $\alpha^2$, and rewrite Eq. \eqref{eq:current_auto_linear} as 
 \begin{equation}
    \mean{I_S} = \frac{q_e}{2\tau_0} \left[ \Delta V_S -v_e \frac{\alpha^2}{e^{\Delta V_S/2}-1}\right].
    \label{eq:av_current_alpha}
\end{equation}
Recalling that the timescale $\tau_0$ is proportional to $v_e$, we see that if we scale the system bias as $\Delta V_S = c v_e$ (for a constant $c>0$), the rectification power at the system side becomes independent of the scale:
\begin{equation}
    T \dot \Sigma_S = \Delta V_S \mean{I_S} \sim -\alpha^2/c \qquad \text{for} \qquad v_e \to 0.
    \label{eq:scaling_system} 
\end{equation}
This result is remarkable since it implies that the mean value $\mean{I_S}$ remains negative in the macroscopic limit, and therefore does not reduce to the positive value obtained from the deterministic treatment. The link between the macroscopic and deterministic limits is broken by the powering of the demon, needed to amplify thermal fluctuations. Indeed, in order to maintain the value of $\alpha^2$ constant, $\Delta V_D$ must be $\Delta V_D = 2\log(1+2\alpha^2/cv_e)$. Then, the current through the demon inverter is $\mean{I_D} \simeq  (q_e/\tau_0)(e^{\Delta V_D/2}-1) = 2q_e\alpha^2/cv_e\tau_0$, and the dissipation power at the demon side scales as 
\begin{equation}
    T\dot \Sigma_D  = \Delta V_D \mean{I_D} \sim \log(v_e^{-1})/v_e^2 \qquad \text{for} \quad v_e \to 0.
    \label{eq:scaling_demon} 
\end{equation}
As a consequence, the thermodynamic efficiency scales as $\eta \sim v_e^{2}/\log(v_e^{-1})$. This scaling is verified numerically by generating stochastic trajectories via the Gillespie algorithm and computing the average currents $\meann{I_{S/D}}$. The results are shown in Figure \ref{fig:efficiency}. We see that a similar scaling is obtained even when the system is working at the critical point $\alpha^2=1$ or in the bistable phase $\alpha^2>1$. Thus, the rectification effects can survive the macroscopic limit if the power available to the MD is appropriately scaled, at the price of a decreasing efficiency.
Finally, the scaling with temperature can be analyzed by recalling that all voltages are expressed in units of $V_T = k_bT/q_e$. Then, from Eq. 
\eqref{eq:av_current_alpha} we see that $\mean{I_S} <0$
only for $V_T > V_T^*$, where $V_T^* = \Delta V_S/\log(1+\alpha^2v_e/\Delta V_S)$ sets the temperature below which the demon stops working.

\emph{Discussion}. 
%
Our results unveil a general mechanism via which the usual correspondence between the macroscopic and deterministic limits can be broken. As one increases the scale of a system, the decreasing thermal fluctuations of macroscopic observables (in this case the voltage of a conductor) can be compensated by increasing the power dedicated to amplify them. In this way, microscopic thermal fluctuations can be transported to the macroscale where they can be employed to produce phenomena (in this case the inversion of a current) that are not allowed by the deterministic dynamics obtained from the macroscopic limit taken at fixed powering.


\emph{Acknowledgments.}
This research was supported by the project INTER/FNRS/20/15074473 funded by F.R.S.-FNRS (Belgium) and FNR (Luxembourg),
and by the FQXi foundation project FQXi-IAF19-05. 

\bibliographystyle{unsrt}
\bibliography{references.bib}

\begin{thebibliography}{10}

\bibitem{leff2002}
Harvey Leff and Andrew~F Rex.
\newblock {\em Maxwell's Demon 2 Entropy, Classical and Quantum Information,
  Computing}.
\newblock CRC Press, 2002.

\bibitem{bennett1982}
Charles~H Bennett.
\newblock The thermodynamics of computation—a review.
\newblock {\em International Journal of Theoretical Physics}, 21(12):905--940,
  1982.

\bibitem{parrondo2015}
Juan~MR Parrondo, Jordan~M Horowitz, and Takahiro Sagawa.
\newblock Thermodynamics of information.
\newblock {\em Nature physics}, 11(2):131--139, 2015.

\bibitem{Sagawa2008}
Takahiro Sagawa and Masahito Ueda.
\newblock {Second Law of Thermodynamics with Discrete Quantum Feedback
  Control}.
\newblock {\em Phys. Rev. Lett.}, 100(8):080403, Feb 2008.

\bibitem{Sagawa2009}
Takahiro Sagawa and Masahito Ueda.
\newblock Minimal energy cost for thermodynamic information processing:
  measurement and information erasure.
\newblock {\em Physical review letters}, 102(25):250602, 2009.

\bibitem{Cao2009}
F.~J. Cao and M.~Feito.
\newblock {Thermodynamics of feedback controlled systems}.
\newblock {\em Phys. Rev. E}, 79(4):041118, Apr 2009.

\bibitem{Sagawa2012}
Takahiro Sagawa and Masahito Ueda.
\newblock {Nonequilibrium thermodynamics of feedback control}.
\newblock {\em Phys. Rev. E}, 85(2):021104, Feb 2012.

\bibitem{esposito2012}
Massimiliano Esposito and Gernot Schaller.
\newblock Stochastic thermodynamics for “maxwell demon” feedbacks.
\newblock {\em EPL (Europhysics Letters)}, 99(3):30003, 2012.

\bibitem{barato2014}
A.~C. Barato and U.~Seifert.
\newblock {Unifying Three Perspectives on Information Processing in Stochastic
  Thermodynamics}.
\newblock {\em Phys. Rev. Lett.}, 112(9):090601, Mar 2014.

\bibitem{horowitz2014}
Jordan~M Horowitz and Massimiliano Esposito.
\newblock Thermodynamics with continuous information flow.
\newblock {\em Physical Review X}, 4(3):031015, 2014.

\bibitem{hartich2014}
David Hartich, Andre~C Barato, and Udo Seifert.
\newblock Stochastic thermodynamics of bipartite systems: transfer entropy
  inequalities and a maxwell’s demon interpretation.
\newblock {\em Journal of Statistical Mechanics: Theory and Experiment},
  2014(2):P02016, 2014.

\bibitem{Serreli2007}
Viviana Serreli, Chin-Fa Lee, Euan~R. Kay, and David~A. Leigh.
\newblock {A molecular information ratchet - Nature}.
\newblock {\em Nature}, 445:523--527, Feb 2007.

\bibitem{Price2008}
Gabriel~N. Price, S.~Travis Bannerman, Kirsten Viering, Edvardas Narevicius,
  and Mark~G. Raizen.
\newblock {Single-Photon Atomic Cooling}.
\newblock {\em Phys. Rev. Lett.}, 100(9):093004, Mar 2008.

\bibitem{Raizen2009}
Mark~G. Raizen.
\newblock {Comprehensive Control of Atomic Motion}.
\newblock {\em Science}, Jun 2009.

\bibitem{Bannerman2009}
S.~Travis Bannerman, Gabriel~N. Price, Kirsten Viering, and Mark~G. Raizen.
\newblock {Single-photon cooling at the limit of trap dynamics: Maxwell's demon
  near maximum efficiency}.
\newblock {\em New J. Phys.}, 11(6):063044, Jun 2009.

\bibitem{Kumar2018}
Aishwarya Kumar, Tsung-Yao Wu, Felipe Giraldo, and David~S. Weiss.
\newblock {Sorting ultracold atoms in a three-dimensional optical lattice in a
  realization of Maxwell{'}s demon - Nature}.
\newblock {\em Nature}, 561:83--87, Sep 2018.

\bibitem{Toyabe2010}
Shoichi Toyabe, Takahiro Sagawa, Masahito Ueda, Eiro Muneyuki, and Masaki Sano.
\newblock {Experimental demonstration of information-to-energy conversion and
  validation of the generalized Jarzynski equality - Nature Physics}.
\newblock {\em Nat. Phys.}, 6:988--992, Dec 2010.

\bibitem{saha2021}
Tushar~K Saha, Joseph~NE Lucero, Jannik Ehrich, David~A Sivak, and John
  Bechhoefer.
\newblock Maximizing power and velocity of an information engine.
\newblock {\em Proceedings of the National Academy of Sciences}, 118(20), 2021.

\bibitem{Koski2014Sep}
Jonne~V. Koski, Ville~F. Maisi, Jukka~P. Pekola, and Dmitri~V. Averin.
\newblock {Experimental realization of a Szilard engine with a single
  electron}.
\newblock {\em Proc. Natl. Acad. Sci. U.S.A.}, 111(38):13786--13789, Sep 2014.

\bibitem{Koski2014Jul}
J.~V. Koski, V.~F. Maisi, T.~Sagawa, and J.~P. Pekola.
\newblock {Experimental Observation of the Role of Mutual Information in the
  Nonequilibrium Dynamics of a Maxwell Demon}.
\newblock {\em Phys. Rev. Lett.}, 113(3):030601, Jul 2014.

\bibitem{Koski2015}
J.~V. Koski, A.~Kutvonen, I.~M. Khaymovich, T.~Ala-Nissila, and J.~P. Pekola.
\newblock {On-Chip Maxwell's Demon as an Information-Powered Refrigerator}.
\newblock {\em Phys. Rev. Lett.}, 115(26):260602, Dec 2015.

\bibitem{Chida2017}
Kensaku Chida, Samarth Desai, Katsuhiko Nishiguchi, and Akira Fujiwara.
\newblock {Power generator driven by Maxwell{'}s demon - Nature
  Communications}.
\newblock {\em Nat. Commun.}, 8(15310):1--7, May 2017.

\bibitem{Camati2016}
Patrice~A. Camati, John P.~S. Peterson, Tiago~B.
  Batalh{\ifmmode\tilde{a}\else\~{a}\fi}o, Kaonan Micadei, Alexandre~M. Souza,
  Roberto~S. Sarthour, Ivan~S. Oliveira, and Roberto~M. Serra.
\newblock {Experimental Rectification of Entropy Production by Maxwell's Demon
  in a Quantum System}.
\newblock {\em Phys. Rev. Lett.}, 117(24):240502, Dec 2016.

\bibitem{Vidrighin2016}
Mihai~D. Vidrighin, Oscar Dahlsten, Marco Barbieri, M.~S. Kim, Vlatko Vedral,
  and Ian~A. Walmsley.
\newblock {Photonic Maxwell's Demon}.
\newblock {\em Phys. Rev. Lett.}, 116(5):050401, Feb 2016.

\bibitem{Cottet2017}
Nathana{\ifmmode\ddot{e}\else\"{e}\fi}l Cottet,
  S{\ifmmode\acute{e}\else\'{e}\fi}bastien Jezouin, Landry Bretheau, Philippe
  Campagne-Ibarcq, Quentin Ficheux, Janet Anders, Alexia
  Auff{\ifmmode\grave{e}\else\`{e}\fi}ves, R{\ifmmode\acute{e}\else\'{e}\fi}mi
  Azouit, Pierre Rouchon, and Benjamin Huard.
\newblock {Observing a quantum Maxwell demon at work}.
\newblock {\em Proc. Natl. Acad. Sci. U.S.A.}, 114(29):7561--7564, Jul 2017.

\bibitem{Masuyama2018}
Y.~Masuyama, K.~Funo, Y.~Murashita, A.~Noguchi, S.~Kono, Y.~Tabuchi,
  R.~Yamazaki, M.~Ueda, and Y.~Nakamura.
\newblock {Information-to-work conversion by Maxwell{'}s demon in a
  superconducting circuit quantum electrodynamical system - Nature
  Communications}.
\newblock {\em Nat. Commun.}, 9(1291):1--6, Mar 2018.

\bibitem{Naghiloo2018}
M.~Naghiloo, J.~J. Alonso, A.~Romito, E.~Lutz, and K.~W. Murch.
\newblock {Information Gain and Loss for a Quantum Maxwell's Demon}.
\newblock {\em Phys. Rev. Lett.}, 121(3):030604, Jul 2018.

\bibitem{Ribezzi-Crivellari2019}
M.~Ribezzi-Crivellari and F.~Ritort.
\newblock {Large work extraction and the Landauer limit in a continuous Maxwell
  demon - Nature Physics}.
\newblock {\em Nat. Phys.}, 15:660--664, Jul 2019.

\bibitem{Najera-Santos2020}
Baldo-Luis Najera-Santos, Patrice~A. Camati, Valentin
  M{\ifmmode\acute{e}\else\'{e}\fi}tillon, Michel Brune, Jean-Michel Raimond,
  Alexia Auff{\ifmmode\grave{e}\else\`{e}\fi}ves, and Igor Dotsenko.
\newblock {Autonomous Maxwell's demon in a cavity QED system}.
\newblock {\em Phys. Rev. Res.}, 2(3):032025, Jul 2020.

\bibitem{pathria2021}
R.~K. Pathria and Paul Beale.
\newblock {\em {Statistical Mechanics}}.
\newblock Academic Press, Cambridge, MA, USA, Feb 2021.

\bibitem{roldan2014}
{\'E}~Rold{\'a}n, Ignacio~A Martinez, Juan~MR Parrondo, and Dmitri Petrov.
\newblock Universal features in the energetics of symmetry breaking.
\newblock {\em Nature Physics}, 10(6):457--461, 2014.

\bibitem{parrondo2001}
Juan~MR Parrondo.
\newblock The szilard engine revisited: Entropy, macroscopic randomness, and
  symmetry breaking phase transitions.
\newblock {\em Chaos: An Interdisciplinary Journal of Nonlinear Science},
  11(3):725--733, 2001.

\bibitem{wang2006}
Alice Wang, Benton~H Calhoun, and Anantha~P Chandrakasan.
\newblock {\em Sub-threshold design for ultra low-power systems}, volume~95.
\newblock Springer, 2006.

\bibitem{sarpeshkar1993}
Rahul Sarpeshkar, Tobias Delbruck, and Carver~A Mead.
\newblock White noise in {MOS} transistors and resistors.
\newblock {\em IEEE Circuits and Devices Magazine}, 9(6):23--29, 1993.

\bibitem{freitas2021}
Nahuel Freitas, Jean-Charles Delvenne, and Massimiliano Esposito.
\newblock {Stochastic Thermodynamics of Nonlinear Electronic Circuits: A
  Realistic Framework for Computing Around $kT$}.
\newblock {\em Phys. Rev. X}, 11(3):031064, Sep 2021.

\bibitem{freitas2021LR}
Nahuel Freitas, Gianmaria Falasco, and Massimiliano Esposito.
\newblock {Linear response in large deviations theory: a method to compute
  non-equilibrium distributions}.
\newblock {\em New J. Phys.}, 23(9):093003, Sep 2021.

\end{thebibliography}

\onecolumngrid
\includepdf[pages={1}, scale=1, openright=true]{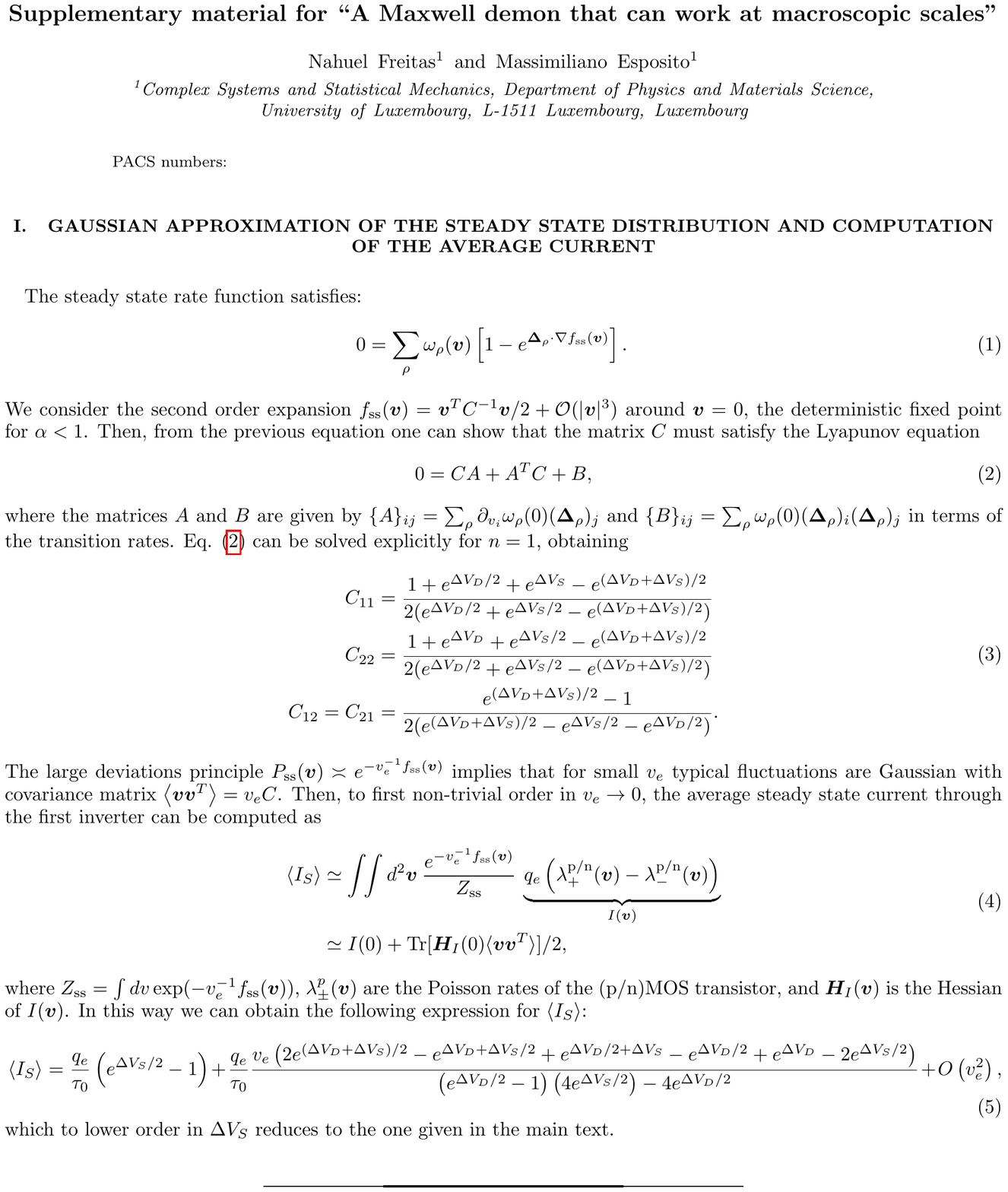}
\thispagestyle{empty}

\end{document}